\documentclass[12pt]{JHEP3}
\title{U(N) spinning particles and
 higher spin equations on complex manifolds}
\author{Fiorenzo Bastianelli and Roberto Bonezzi\\ Dipartimento di Fisica, Universit\`{a} di Bologna and INFN, Sezione di Bologna\\ via Irnerio 46, I-40126 Bologna, Italy\\ E-mail:
\email{bastianelli@bo.infn.it, bonezzi@bo.infn.it}}
\abstract{
Guided by a spinning particle model with $U(N)$-extended supergravity on the worldline
we derive higher spin equations on complex manifolds.
Their minimal formulation is in term of gauge fields which satisfy suitable constraints.
The latter can be relaxed by introducing compensator fields.
There is an obstruction to define these systems on arbitrarily curved spaces,
just as in the usual theory of higher spin fields, but we show how to couple them
to K\"ahler manifolds of constant holomorphic curvature.
Quite interestingly, the first class gauge algebra defining the $U(N)$ particles
on these manifolds is quadratic and realizes the zero mode sector
of certain nonlinear $U(N)$ superconformal algebras introduced
sometimes ago by Bershadsky and Knizhnik in 2D.}

\usepackage[latin1]{inputenc}
\usepackage{indentfirst}
\usepackage{graphicx}
\usepackage{amssymb}
\usepackage{amsmath}
\usepackage{latexsym}
\usepackage{amsthm}
\usepackage{youngtab}
\newcommand{\de}{\partial} 
\newcommand{\braket}[2]{ \langle #1 \lvert #2 \rangle }  
\newcommand{\ket}[1]{\lvert #1 \rangle} 
\newcommand{\ass}[1]{\lvert #1 \rvert} 

\newcommand{\be}{\begin{equation}}
\newcommand{\ee}{\end{equation}}
\newcommand{\bea}{\begin{eqnarray}}
\newcommand{\eea}{\end{eqnarray}}

\begin{document}

\section{Introduction}

Spinning particle models are quite useful for describing field theories in first
quantization. In particular, spinning particles with gauged $O(N)$-extended susy on the
worldline \cite{Gershun:1979fb,Howe:1988ft} can be used to describe properties of fields
of spin $N/2$ \cite{Siegel:1999ew,Bastianelli:2007pv,Bastianelli:2008nm}.

In this paper we analyze spinning particles with gauged $U(N)$-extended
susy on the worldline and use them to derive gauge invariant higher spin equations
on certain complex manifolds. The  $U(N)$ particles for
$N=1,2$ were originally introduced in \cite{Marcus:1994em}
as a dimensional reduction of the $N=2$ string, and generalized
to arbitrary  $N$ in \cite{Marcus:1994mm}. Their Dirac quantization
introduces constraints on the particle Hilbert space
that are interpreted as equations of motion for certain tensor fields
with holomorphic indices satisfying the symmetries of a rectangular
Young tableau \cite{Marcus:1994mm}.
We analyze these equations on the flat complex space $\mathbb{C}^d$.
By integrating a subset of them
in terms of gauge potentials we are led to gauge invariant field equations which are
quite  similar in form to the higher spin equations introduced by Fronsdal \cite{Fronsdal:1978rb}.

An example of these equations is that of a gauge field
$\varphi_{\mu_1 \ldots \mu_N}$ with $N$ symmetric holomorphic indices
(we use complex coordinates $x^\mu, \bar x^{\bar \mu}$ of $\mathbb{C}^d$; tensor indices
are raised and lowered with the flat hermitian metric $\delta_{\mu\bar \nu}$).
It satisfies the equation
\begin{equation}
\partial_\alpha \bar\partial^\alpha
\varphi_{\mu_1 \ldots \mu_N}
-\sum_{i=1}^N \partial_{\mu_i}  \bar \partial^\alpha \varphi_{\mu_1 .. \alpha .. \mu_N}
=0
\end{equation}
where the $\alpha$ index in the second term is located in $i$-th position.
The gauge invariance is given by
\begin{equation}
\delta \varphi_{\mu_1 \ldots \mu_N} =
\partial_{\mu_1} \lambda_{\mu_2 ... \mu_N} + \ cyclic\ perm.
\end{equation}
where the gauge parameter  $\lambda_{\mu_2 ... \mu_N}$ has $N-1$ symmetric
holomorphic indices and is
constrained by $\bar \partial^\alpha \lambda_{\alpha \mu_3 ... \mu_N} =0$.
For consistency the gauge field must also satisfy a differential constraint
$\bar \partial^{\alpha} \bar \partial^{\beta}
\varphi_{\alpha \beta \mu_3 ... \mu_N} =0$.
These equations are very much reminiscent of Fronsdal's equations. Since
there is no invariant concept of taking traces on holomorphic indices, the usual trace constraints that appear in Fronsdal's formulation
are naturally substituted here by differential constraints.

The constraints on gauge fields and on gauge parameters can be relaxed
by adding compensator fields.
For example in the above case with  $N=2$, one can introduce a single
compensator field $\rho$ and the equation reads
\begin{equation}
\partial_\alpha \bar\partial^\alpha \varphi_{\mu\nu}
-\partial_{\mu} \bar \partial^\alpha \varphi_{\alpha\nu}
-\partial_{\nu} \bar \partial^\alpha \varphi_{\mu\alpha}
=\partial_{\mu} \partial_{\nu}\rho
\end{equation}
with gauge symmetry
\begin{equation}
\delta \varphi_{\mu \nu} = \partial_{\mu} \lambda_{\nu} +\partial_{\nu} \lambda_{\mu}
\ , \qquad
\delta \rho = -2 \bar\partial^\alpha\lambda_{\alpha} \ .
\end{equation}
This is reminiscent of the Francia-Sagnotti construction
\cite{Francia:2002aa1} for relaxing the trace constraints of standard
higher spin gauge theories using compensator fields.

We derive equations also for more general tensor fields
with the symmetry type of a rectangular Young tableaux with $p$ rows and $N$ columns.
We do so by using the compact notation provided by the quantum mechanical
operators of the spinning particle.

The equations just discussed are defined on a flat complex space,
viewed as a K\"ahler manifold, but it is interesting to study
if they can be extended to more general K\"ahler spaces.
While it is known that the $U(N)$ particles for
$N=1,2$ can be coupled to any K\"ahler manifold \cite{Marcus:1994em},
it was thought that for $N>2$  these
particles could only be consistent on flat  manifolds,
as the standard susy transformation rules do not leave the particle action invariant
on a curved space \cite{Marcus:1994mm}.
We can actually show that a coupling is still possible for
K\"ahler manifolds with constant holomorphic curvature.
To achieve this result we use a hamiltonian approach and notice
that the first class algebra defining the model
closes on K\"ahler manifolds with constant holomorphic curvature,
though in a nonlinear way. In fact we obtain a quadratic first class
algebra that, quite interestingly, is seen to coincide with the zero mode sector
of two dimensional nonlinear $U(N)$ superconformal algebras, introduced sometimes ago by
Bershadsky and Knizhnik  \cite{Bershadsky:1986ms,Knizhnik:1986wc}.
This result is consistent with \cite{Marcus:1994mm}
in that the susy transformations rules associated to a nonlinear algebra
differ from the one employed in \cite{Marcus:1994mm}.
The corresponding gauge invariant differential equations can similarly be defined on
such complex spaces.

Having understood that $U(N)$ particles and related gauge invariant field equations
can be defined on a non trivial class of curved spaces, it is interesting to study
their quantum properties. We begin this analysis using a first quantized
path integral description. This worldline approach
is quite flexible and efficient, and by using closed worldlines one can study directly
the one loop effective action associated to the field equations
described above. To construct the path integral is necessary to gauge-fix
the particle action and identify the correct measure over the
moduli space of inequivalent gauge choices. We start considering a flat
target space and compute the physical degrees of freedom. This gives a check
on the path integral measure, which can be used to compute more general observables.

While the complex nature of target space does not suggest us an immediate physical
application of these higher spin equations (either the target space has an even number
of times, or no time direction at all) they might still be useful to describe properties
of complex manifolds or for developing additional intuition on the standard theory of
higher spin fields (see \cite{Sorokin:2004ie} for reviews).
From this point of view it would be quite
interesting to search for consistent nonlinear extensions of the free equations
described here.

We structure our paper as follows.
We review the $U(N)$ spinning particle in section 2 using a hamiltonian formulation.
Dirac quantization is analyzed in section 3. In particular
we describe in this section
how the constraints (``curvature formulation'')
can be partially integrated to produce
the equations of motion for gauge fields introduced above (``gauge field formulation'').
In section 4 we prove the consistency of the coupling to K\"ahler manifolds of
constant holomorphic curvature.
In section 5 we construct the worldline path integral  and compute
the number of physical degrees of freedom. Sections 6 contains our conclusions
and an outlook.

\section{The $U(N)$ spinning particle in flat space}\label{section2}

We consider an even dimensional flat space, viewed as the flat K\"ahler manifold
$\mathbb{C}^d$,
with $D=2d$ real dimensions; the bosonic fields $x^M(\tau)$, interpreted as
space-time coordinates, split into complex
components $x^\mu(\tau)$ and $\bar{x}^{\bar{\mu}}(\tau)$,
with $\mu=1\ldots d$. They are paired
with fermionic superpartners $\psi^\mu_i(\tau)$ and $\bar{\psi}^{\bar{\mu}i}(\tau)$, $i=1\ldots
N$, belonging to the $\mathbf{N}$ and $\mathbf{\bar{N}}$ of $U(N)$, respectively. The flat metric
in complex coordinates is simply $\delta_{\mu\bar{\nu}}$, the other components being zero.
With these ingredients the phase space action
\begin{equation}
S=\int_0^1d\tau\,\big[p_\mu\dot{x}^\mu+\bar{p}_{\bar{\mu}}\dot{\bar{x}}^{\bar{\mu}}
+i\bar{\psi}^i_\mu\dot{\psi}^\mu_i
-p_\mu\bar{p}^\mu
\big]
\end{equation}
describes the motion of a free particle with a pseudoclassical spin associated
to the Grassmann coordinates.
This system  enjoys various conserved quantities, including those
corresponding to the $U(N)$-extended supersymmetry on the worldline
\begin{equation}
H=p_\mu \bar{p}^\mu \ , \quad
Q_i=\psi^\mu_i p_\mu  \ , \quad
\bar{Q}^i=\bar{\psi}^{\bar{\mu} i}\bar{p}_{\bar{\mu}} \ , \quad
J_i^j=\psi^\mu_i\bar{\psi}_\mu^j
\end{equation}
where indices
are lowered and raised using the $\delta_{\mu\bar \nu}$ metric and its inverse.
We have chosen normalizations so that $H$ is real, $(Q_i)^* =\bar Q^i$, and
$(J_i^j)^* = J^i_j$, so that $J_i^i$ is real for any fixed $i$.
The fundamental Poisson brackets are easily read off from the symplectic term of
the action
\begin{equation}
\{x^\mu, p_\nu \}_{_{PB}}=\delta^\mu_\nu \ , \quad
\{\bar x^{\bar \mu}, \bar p_{\bar \nu} \}_{_{PB}}=\delta^{\bar\mu}_{\bar \nu}\ , \quad
\{\psi_i^\mu,\bar\psi^{\bar \nu j} \}_{_{PB}} = -i \delta^{\mu\bar \nu} \delta^j_i
\label{pb}
\end{equation}
and the above conserved charges generate symmetry transformations
through Poisson brackets (using $\delta z=\{z,{\cal G}\}_{_{PB}}$ with
${\cal G}\equiv \xi H + i\bar \epsilon^i Q_i + i \epsilon_i \bar Q^i + \alpha_i^j J_j^i$)
\begin{equation}
\begin{split}
\delta x^\mu &= \xi\bar{p}^\mu+i\bar{\epsilon}^i\psi^\mu_i \ ,\qquad\quad \ \delta
\bar{x}^{\bar{\mu}}=\xi p^{\bar{\mu}}+i\epsilon_i\bar{\psi}^{\bar{\mu} i}\\
\delta\psi^\mu_i &= -\epsilon_i\bar{p}^\mu+i\alpha^j_i\psi^\mu_j\ ,\qquad
\delta\bar{\psi}^{\bar{\mu}i}=-\bar{\epsilon}^ip^{\bar{\mu}}
-i\alpha^i_j\bar{\psi}^{\bar{\mu}j}\\
\delta p_\mu &= 0\ ,\qquad \qquad\qquad \quad \ \ \,
\delta\bar{p}_{\bar{\mu}}=0 \ ,
\end{split}
\label{transf}
\end{equation}
which correspond to rigid time translations with parameter $\xi$,
$N$ complex supersymmetries with grassmannian parameters $\epsilon_i$ and
$\bar{\epsilon}^i$, and $U(N)$ rotations parametrized by $\alpha^i_j$.
The explicit $U(N)$-extended supersymmetry
algebra is easily computed
\begin{equation}\label{classical constraints}
\begin{split}
\{Q_i,\bar Q^j\}_{_{PB}} &= -i \delta_i^j H \\
\{J_i^j,Q_k\}_{_{PB}} &= -i \delta^j_k Q_i  \ ,
\qquad \ \
\{J_i^j,\bar Q^k \}_{_{PB}} = i\delta_i^k \bar Q^j
 \\
 \{J_i^j,J_k^l \}_{_{PB}} &= i\delta_i^l J^j_k -i\delta^j_k J_i^l
\end{split}
\end{equation}
with other independent Poisson brackets vanishing.

The model we are interested in is obtained by gauging this first class algebra
through the  introduction of corresponding gauge fields:
an einbein $e(\tau)$ for time translations, complex gravitini
$\chi_i(\tau)$ and $\bar{\chi}^i(\tau)$ for the extended supersymmetry, and a $U(N)$ gauge field $a^i_j(\tau)$ for the rotations.
These fields correspond to the gauge fields of a $U(N)$-extended supergravity
on the worldline, and the full action of the $U(N)$ spinning particle becomes
\begin{equation}\label{Sinvariant}
S=\int_0^1d\tau\,\big[p_\mu\dot{x}^\mu+\bar{p}_{\bar{\mu}}\dot{\bar{x}}^{\bar{\mu}}
+i\bar{\psi}^i_\mu\dot{\psi}^\mu_i-e\underbrace{p_\mu\bar{p}^\mu}_{H}
-i\bar{\chi}^i \underbrace{p_\mu\psi^\mu_i}_{Q_i}
-i\chi_i \underbrace{\bar{p}_{\bar{\mu}}\bar{\psi}^{\bar{\mu}\,i}}_{\bar Q^i}
-a_j^i (\underbrace{\psi^\mu_i\bar{\psi}_\mu^j}_{J_i^j} -s\delta^j_i)\big]
\end{equation}
where we have inserted also a Chern-Simons coupling $s$ for the $U(1)$ part
of the gauge group $U(N)$, since it is invariant by itself.
The supergravity gauge fields
turn the rigid symmetries of eqs. (\ref{transf}) into local ones
and transform as follows
\begin{equation}\label{gaugetransf}
\begin{split}
\delta e &= \dot{\xi}+i\bar{\chi}^i\epsilon_i+i\chi_i\bar{\epsilon}^i\\
\delta\chi_i &=
\dot{\epsilon}_i-ia^k_i\epsilon_k+i\alpha^k_i\chi_k=\mathcal{D}\epsilon_i+i\alpha^k_i\chi_k\\
\delta\bar{\chi}^i &=
\dot{\bar{\epsilon}}^i+ia^i_k\bar{\epsilon}^k-i\alpha^i_k\bar{\chi}^k
=\mathcal{D}\bar{\epsilon}^i-i\alpha^i_{\,k}\bar{\chi}^k\\
\delta a^i_j &=
\dot{\alpha}^i_j -i a_j^k\alpha_k^i+ia^i_k\alpha^k_j=\mathcal{D}\alpha^i_j
\end{split}
\end{equation}
where $\mathcal{D}$ stands for the $U(N)$ covariant derivative.\\
From the phase space action \eqref{Sinvariant} it is immediate to see that the equations of motion of
the gauge fields $G\equiv (e,\chi,\bar{\chi},a)$
constrain the Noether charges to vanish
\begin{equation}
\frac{\delta S}{\delta G}=0\qquad \Rightarrow\qquad H=Q_i=\bar{Q}^i=J_i^j-s\delta_i^j=0\;.
\end{equation}
The  Poisson brackets of these generators form the $U(N)$-extended supersymmetry algebra that, as we shall see, ceases to be first class for $N>2$ on arbitrary curved manifolds. This hints to a
fundamental obstruction in imposing the constraints listed above and is the signal, from a
worldline point of view, of the difficulties that arise in coupling higher spin particles to curved spaces. We will discuss this issue in more depth in section 4.

Eliminating the momenta $p$ and $\bar p$ one obtains the action in configuration space
\begin{equation}\label{Sinvariantconfspace}
S[X,G]=\int_0^1d\tau\big[e^{-1}\big(\dot{x}^\mu-i\bar{\chi}^i\psi_i^\mu\big)
\big(\dot{\bar{x}}_\mu-i\chi_j\bar{\psi}_\mu^j\big)+i\bar{\psi}_\mu^i\big(\delta_i^j\de_\tau
-ia_i^j\big)\psi_j^\mu +s a_i^i\big]
\end{equation}
where $X\equiv (x,\bar{x}, \psi,\bar{\psi})$ and $G\equiv (e,\chi,\bar{\chi},a)$.
We shall use this form when constructing the path integral in section 5.

\section{Equations of motion in flat space}\label{section3}

We now use canonical quantization and obtain the
equations of motion in flat space.
From the constraint $H=0$, we see that the system has a constant $\tau$ evolution.
The dynamics of the particle is then entirely contained in
the constraints $H=Q_i=\bar{Q}^i=J_i^j -s\delta_i^j=0$: these
classical statements translate, in the quantum theory, into the selection of the
physical Hilbert space, which is obtained by requiring the symmetry generators to
annihilate physical states, \emph{i.e.}
\begin{equation}\label{constraints}
\ket{\Phi}\in\mathcal{H}_{phys}\quad \Longleftrightarrow \quad
T_a\ket{\Phi}=0\;, \ \ \;T_a=(H\,,\,Q_i\,,\,\bar{Q}^i\,,\,J_i^j -s \delta_i^j)
\end{equation}
where the generators $T_a$ are now to be understood as operators.
The Chern-Simons coupling $s$ will satisfy a quantization condition that can be stated
precisely once
a prescription for resolving the ordering ambiguities contained
in $J_i^j$ is taken care of.
What we have just described is the Dirac quantization procedure, which generalizes
the quantization \`{a} la Gupta-Bleuler of electrodynamics.
As already discussed in \cite{Marcus:1994mm}, the particle
states can be represented by generalized field strengths of the form
$F_{\mu_1^1..\,\mu_m^1,\ldots,\;\mu_1^N..\,\mu_m^N}$, where the integer $m$ is related
to the quantized Chern-Simons coupling $s$.
In particular, the $J$ constraints require that
$F$ is antisymmetric within each block of $m$ indices, symmetric in exchanging
entire blocks, and
in addition
satisfies algebraic Bianchi identities, \emph{i.e.} it belongs to an irreducible representation of $U(d)$ with rectangular $m\times N$ Young tableau:
\begin{equation}\label{Young Tab F}
F_{\mu_1^1..\,\mu_m^1\,,\ldots,\,\mu_1^N..\,\mu_m^N}
\quad \sim \quad
m\;\quad\underbrace{
\hspace{-12pt}\left\{\begin{array}{c}\\[-15pt]\yng(4,4,4)\end{array}\right.}_{N} \ .
\end{equation}
The $Q$ and $\bar{Q}$ constraints enforce
generalized Maxwell equations, while the $H$ constraint is automatically satisfied
in virtue of the constraint algebra.

We now proceed in deriving the results stated above: looking at the fundamental
(anti)-commutation relations, which follows from the classical Poisson brackets (\ref{pb}),
\begin{equation}\label{fundamentalcommutators}
\big[x^\mu\,,\,p_\nu\big]=i\delta^\mu_\nu\ ,
\quad \big[\bar{x}^{\bar{\mu}}\,,\,\bar{p}_{\bar{\nu}}\big]
=i\delta^{\bar{\mu}}_{\bar{\nu}}\ , \quad
\big\{\psi^\mu_i\,,\,\bar{\psi}^{\bar{\nu}\,j}\big\}=\delta^{\mu\bar{\nu}}\delta_i^j\ ,
\end{equation}
one can decide to project the states of the Hilbert space onto the
$x^\mu$, $\bar{x}^{\bar{\mu}}$ and $\psi^\mu_i$ eigenstates. In this way $x$, $\bar{x}$ and $\psi$
act by multiplication, while momenta $p$, $\bar p$ and $\bar{\psi}$ act as derivatives: $p_\mu\sim -i\de_\mu$,
$\bar{p}_{\bar{\mu}}\sim -i\bar \de_{\bar{\mu}}$ and
$\bar{\psi}^{\bar{\mu}i}\sim\frac{\de}{\de\psi^\mu_i}$. The states are thus represented by
functions of $x$, $\bar{x}$ and $\psi$: $\ket{F}\sim\braket{x,\bar{x},\psi}{F}=F(x,\bar{x},\psi)$.
With this realization the symmetry generators $T_a$ read
\begin{equation}\label{represent of symmetry generators}
\begin{split}
J_i^j -s \delta_i^j&= \psi_i\cdot\frac{\de}{\de\psi_j}-m\,\delta_i^j \\[1mm]
Q_i &= -i\psi_i^\mu\,\de_\mu \\[1mm]
\bar{Q}^i &=-i\,\frac{\de}{\de\psi^\mu_i}\,\bar\de^\mu \\[1mm]
H &= -\delta^{\mu\bar{\nu}}\,\de_\mu\bar\de_{\bar{\nu}}
\end{split}
\end{equation}
where $\bar\de^\nu=\delta^{\nu\bar{\mu}}\,\bar\de_{\bar{\mu}}$.
Ordering ambiguities are only present in the $J$ constraint. We have resolved them
by using a graded-symmetric ordering,
which coincides with the natural regularization that arises form the
path  integral of section 5,
\begin{equation}\label{3.5}
J_i^j=\frac12 (\psi_i^\mu \bar \psi_\mu^j -\bar \psi_\mu^j\psi_i^\mu)
=\psi_i^\mu \bar \psi_\mu^j -\frac{d}{2} \delta_i^j
\quad \Longrightarrow \quad J_i^j -s \delta_i^j =
\psi_i\cdot\frac{\de}{\de\psi_j}-m\,\delta_i^j
\end{equation}
where we have set $m\equiv (\frac{d}{2} + s)$. The quantum constraints
satisfy a first class algebra corresponding to the quantum version
of (\ref{classical constraints})
\begin{equation}\label{quantum constraints}
\begin{split}
\big\{Q_i,\bar{Q}^j\big\}&=\delta_i^j H \\
\big[J_i^j,Q_k\big]&=\delta^j_k Q_i\ , \qquad\ \ \ \
\big[J_i^j,\bar{Q}^k\big]=-\delta_i^k \bar{Q}^j\\
\big[J_i^j,J_k^l\big]&=\delta^j_k J_i^l-\delta^l_i J^j_k
\end{split}
\end{equation}
while other independent graded-commutators vanish.
Here we have used the simple $J_i^j$ generators, but it is evident that the same result
holds by substituting them with $J_i^j-s\delta_i^j$ since the Chern-Simons term is central
and in addition it cancels on right hand sides.\\
Due to the grassmannian nature of the
$\psi$ variables, the states have a finite Taylor expansion in $\psi$'s
\begin{equation}\label{general expansion}
\ket{F}\sim\sum_{A_i=0}^d
F_{\mu_1^1..\,\mu_{A_1}^1,\ldots,\,\mu_1^N ..\,\mu_{A_N}^N}(x,\bar{x})
\,\psi_1^{\mu_1^1}..\,\psi_1^{\mu_{A_1}^1}\ldots\,\psi_N^{\mu_1^N}..\,\psi_N^{\mu_{A_N}^N}
\end{equation}
and we can now study which of them survive the constraint equations.

First we consider the $J_i^j$ constraints.  The $J_i^i$ constraint
at fixed $i$ counts fermions of $i$-th type
and fixes them to be $m$ in number, see \eqref{3.5}.
Thus $m$ must be an integer and this, in turn,
fixes the possible quantized values of the Chern-Simons coupling $s$.
Hence, the only term of \eqref{general expansion} surviving this constraint is
\begin{equation}\label{expansion}
F_{\mu_1^1..\,\mu_m^1,\ldots,\,\mu_1^N..\,\mu_m^N}\
\psi_1^{\mu_1^1}..\psi_1^{\mu_{m}^1}
\ldots\psi_N^{\mu_1^N}..\psi_N^{\mu_{m}^N}
\end{equation}
\emph{i.e.} a tensor with $N$ blocks of $m$ indices.
In term of complex geometry, the tensor
$F_{\mu_1^1..\mu_{m}^1,...,\mu_1^N..\mu_{m}^N}(x,\bar{x})$
can be thought of a
differential multiple $(m,0)$-form: in fact each $\psi_i$ block in
\eqref{expansion} plays the role of a basis
for the $(m,0)$-forms, $dx^{\mu_1}\wedge...\wedge dx^{\mu_{m}}$.
The $J_i^j$ constraint for $i\neq j$ then ensures algebraic Bianchi identities:
it picks an index of the $j$-th block, antisymmetrizes it with
those of the $i$-th block, and set the resulting tensor to zero. For example, the
$J_1^2$  constraint gives
\begin{equation}
F_{[\mu_1^1..\,\mu^1_{m},\,\mu^2_1]...,\,\mu_1^N..\,\mu_{m}^N}=0
\end{equation}
and so on. As a consequence, the tensor
$F_{\mu_1^1..\,\mu_m^1,\ldots,\,\mu_1^N..\,\mu_m^N}$
has $N$ blocks of $m$ antisymmetric indices and is symmetric under exchanges
of blocks. The antisymmetry within each block
is evident from the Grassmann nature of the $\psi$'s, while symmetry between blocks
can be understood considering particular $U(N)$ transformations.
In fact, a $\frac{\pi}{2}$
rotation in the $i-j$ plane sends $\psi_i$ in $\psi_j$ and $\psi_j$ in $-\psi_i$.
The final effect of this $U(N)$ transformation is to exchange the $i$-th  and
$j$-th blocks of indices on the tensor $F$ in \eqref{expansion}
without any additional sign.
Since this is a $U(N)$ transformation connected to the identity,
it can be cast in the form $e^{i \alpha^i_j J_i^j}$ for some $\alpha_j^i$
with $i\neq j$.
Requiring $ J_i^j\ket{F}=0$ produces the anticipated
symmetry between the $i$-th and $j$-th blocks of indices of the tensor $F$.
All these algebraic symmetries are summarized by saying that $F$
belongs to an irreducible representation  of the group $U(d)$
described by the Young tableau in eq. \eqref{Young Tab F}.
Finally,
using the representation \eqref{represent of symmetry generators}
of the operators $Q_i$ and $\bar{Q}^i$, it is straightforward to see that
their constraints impose the following generalized Maxwell equations on the curvature
$F$
\begin{equation}\label{Bianchi}
\de_{[\mu}F_{\mu_1^1..\,\mu_{m}^1]\,,...,\,\mu_1^N..\,\mu_{m}^N}=0\;,
\end{equation}
\begin{equation}\label{Maxwell}
\bar\de^\mu F_{\mu..\,\mu_{m}^1,...,\,\mu_1^N..\,\mu_{m}^N}=0\;.
\end{equation}

\subsection{Gauge fields}

In analogy with Maxwell, Yang-Mills and higher spin gauge theories, we first try to solve
eq. \eqref{Bianchi}. This equation can be interpreted as an integrability condition.
In the absence of topological obstructions,
the closure of a form $F$ is achieved expressing it as the exterior derivative of a gauge field:
$F=d\phi\,\rightarrow\,dF=0$. In our context we are dealing with
$N$-multiple $(m,0)$-forms, and
we are going to show that \eqref{Bianchi},
that is to say $Q_i\ket{F}=0$, can be solved writing $F$ as the multiple action
(one for each block of indices) of the holomorphic Dolbeault operator $\de$, that sends
forms of bidegree
$(p,q)$ into $(p+1,q)$-forms. As the $\de_{(i)}$ operator\footnote{The index $i$ refers
to the block on which the Dolbeault operator $\de$ acts,
while other blocks are treated as spectators.}
in our quantum mechanical notation is simply $Q_i$, it is useful to define
\begin{equation}
q=Q_1Q_2...\,Q_N
\end{equation}
which is identically annihilated by the $Q_i$'s: $qQ_i=Q_iq=0$, due to $Q_i^2=0$ and to
the fact that $q$ contains already all of the $Q_i$'s. Setting
\begin{equation}\label{F=q phi}
\ket{F}=q\ket{\phi}
\end{equation}
automatically satisfies the $Q$ constraints and, writing down \eqref{F=q phi} in components, we see that $F \sim \de_{(1)}...\,\de_{(N)}\phi$,
where each Dolbeault operator antisymmetrizes only over the
corresponding block of indices.
To solve the $J$ constraints one can take
 $\phi$ to be a
$N$-multiple $(p,0)$-form with $p\equiv m-1$
that forms a $U(d)$ irreducible tensor (a rectangular $p \times N$ Young tableau)
\begin{equation}\label{phi}
\ket{\phi}
\quad \sim\quad
\phi_{\mu_1^1..\,\mu_p^1,...,\,\mu_1^N..\,\mu_p^N}(x,\bar{x})
\quad \sim \quad
p\;\quad\underbrace{
\hspace{-12pt}\left\{\begin{array}{c}\\[-15pt]\yng(4,4)\end{array}\right.}_{N}\;.
\end{equation}
In fact,  we note that
$J_i^j\,q=q\,(J_i^j+\delta_i^j)$.
Thus $(J_i^j-s\delta_i^j)\ket{F}=0$ is satisfied if one requires
\begin{equation}\label{J constraint on phi}
\big(J_i^j-(s-1)\delta_i^j\big)\,\ket{\phi}=0\;,
\end{equation}
that is, $N_i=m-1\equiv p$ if taking $i=j$
(by $N_i\equiv \psi_i\cdot\frac{\de}{\de\psi_i}$ at fixed $i$ we indicate the
number operator that counts the fermions of the $i$-th type), while the
off diagonal equations are the same as for $F$: they impose algebraic Bianchi identities and,
in particular, symmetry between block exchanges.\\
Next it remains to implement the last independent constraint,
$\bar{Q}^i\ket{F}=0$. This produces generalized Maxwell equations for the gauge field.
From \eqref{F=q phi} it is clear that $\bar{Q}^i\,q\ket{\phi}=0$ is
an higher derivative equation of motion for the gauge potential, precisely of order $N+1$.
It is
convenient to use some $Q,\bar{Q}$ algebra in order to factorize from the operator $\bar{Q}^i\,q$ a second order differential operator $G$,
that will play a role analogous to the Fronsdal-Labastida operator
\cite{Fronsdal:1978rb,Labastida:1986ft}
for higher spin fields. Iterated use of $\{Q_i, Q_j\}=0$ and
$\{Q_i,\bar{Q}^j\}=\delta_i^j\,H$ gives (in the following equation $j$ is fixed, not summed)
\begin{equation*}
\begin{split}
\bar{Q}^j\,q &=
\bar{Q}^j\,Q_1Q_2..Q_N=(-1)^{j-1}Q_1..\bar{Q}^jQ_j..Q_N\\
 &=
 (-1)^{j-1}\big(Q_1..Q_{j-1}Q_{j+1}..Q_N\big)\bar{Q}^jQ_j\\
 &= (-1)^{j-1}\big(Q_1..Q_{j-1}Q_{j+1}..Q_N\big)\big(H-Q_j\bar{Q}^j\big)\;.
\end{split}
\end{equation*}
At this point is possible to sum over $j$ in $H-Q_j\bar{Q}^j$, since
the extra terms vanish anyhow,
and cast the equation of motion in the form
\begin{equation}\label{higher derivative eom}
\bar{Q}^jq\ket{\phi}=q^{j}G\ket{\phi}=0
\end{equation}
where, in an obvious notation, $q^{j}\equiv(-1)^jQ_1..Q_{j-1}Q_{j+1}..Q_N$.
$G$ is the second
order operator we were looking for, analogous to the Fronsdal-Labastida operator
without the trace term
\begin{equation}\label{Fronsdal-L operator}
G=-H+Q_i\bar{Q}^i\ \ \sim\ \ \partial_\alpha \bar\partial^\alpha
-\psi_i^\alpha\,\frac{\de}{\de\psi_i^\beta}\,\de_\alpha\bar\de^\beta\;.
\end{equation}
To obtain a second order equation of motion from \eqref{higher derivative eom}
it is necessary
to eliminate the operator $q^{j}$. One way to do this is recalling that a generic expression
containing two $Q$'s represents the kernel of $q^{j}$, that is $q^{j}Q_kQ_l\equiv 0$, and so
a general solution of $q^{i}\big(G\ket{\phi}\big)=0$ is
\begin{equation}\label{Fronsdal Labastida with compensators}
G\ket{\phi}=Q_iQ_j\ket{\rho^{ij}}
\end{equation}
where  $\ket{\rho^{ij}}$ are the compensator fields.
One can present the compensators also in the form
$\ket{\rho^{ij}}=\bar{V}^i\bar{V}^j\ket{\rho}$.
This second form
of writing the compensators is slightly more convenient. Here $\bar{V}^i
\equiv V^\mu\bar{\psi}_\mu^i$
depends on an arbitrary vector field $V^\mu$, and $\ket{\rho}$ is a state that must satisfy
$(J_i^j-(s-1)\delta_i^j)\ket{\rho}=0$ (because of eq. \eqref{J constraint on phi} and $[G,\,J_i^j]=0$)
and thus is represented by a tensor with the same structure and Young tableau of $\phi$. The action
of $\bar{V}^i$ is to eliminate one $\psi$ from the $i$-th block and saturate the corresponding
index of the $\rho$ tensor with $V^\mu$. Therefore the compensator  $\rho^{ij}$ has $N-2$ blocks
with $p$ antisymmetric indices and two blocks, the $i$-th and $j$-th ones,
with $p-1$ indices. Its Young tableau has the form
\begin{equation}\label{Young Tab rho}
\rho^{ij}\quad \sim\quad p\;\quad\underbrace{
\hspace{-12pt}\left\{\begin{array}{c}\\[-15pt]\yng(4,4,2)\end{array}\right.}_{N}\;.
\end{equation}
The key feature of eq. \eqref{Fronsdal Labastida with compensators}
is to be a second order wave
equation. The price for this is the introduction of the auxiliary
fields $\rho^{ij}$. Of course one would like also
to obtain an equation without compensators, $G\ket{\phi}=0$.
This is indeed possible using gauge symmetries. In fact,
in theories where the physical field strength is expressed in terms
of a potential, one expects the presence of a gauge symmetry.\\

\indent \textbf{Gauge symmetry}\\
\noindent In term of forms if $F=d\phi$, the gauge transformation leaving $F$ invariant is
$\delta\phi=d\Lambda$. In our model the gauge symmetry enjoyed by the curvature $F$
is an ``higher
spin'' generalization of the linearized diffeomorphisms of general relativity, like the gauge
transformations of standard higher spin fields. In our operator formalism, exterior holomorphic derivatives acting on the $i$-th block are
represented by the supercharge $Q_i$. Thus, recalling that $\ket{F}=q\ket{\phi}$ and
$q Q_i=0$, one finds immediately an invariant way of writing down the gauge
transformations that leave the $F$ tensor invariant
\begin{equation}\label{gauge transformation phi}
\delta\ket{\phi}=Q_i\ket{\Lambda^i}
\end{equation}
where $\ket{\Lambda^i}$ are the gauge parameters. Again
a slightly more convenient way of writing the gauge parameters is in the form
$\ket{\Lambda^i} =\bar{W}^i\ket{\Lambda}$, where
$\bar{W}^i \equiv W^\mu\bar{\psi}_\mu^i$ with
$W^\mu$ a vector field
and $\ket{\Lambda}$  a state containing
a tensor with the same index structure and
Young tableau of $\ket{\phi}$. These gauge transformations clearly do not affect
$\ket{F}=q\ket{\phi}$, but let us compute how the left hand side of
\eqref{Fronsdal Labastida with compensators} transforms.
Making use of the $Q$, $\bar{Q}$ algebra the gauge variation can be written as
\begin{equation}
G\delta\ket{\phi}=-Q_iQ_j\big(\bar{Q}^i\ket{\Lambda^j}\big)\;,
\end{equation}
and if we want the equations of motion to be gauge invariant, the compensator field
(from this its name) has to cancel the above expression and transform as
\begin{equation}\label{gauge transformation rho}
\delta\ket{\rho^{ij}}=-\bar{Q}^{[i}\ket{\Lambda^{j]}}\;.
\end{equation}

It is well known from higher spin field theories
\cite{Francia:2002aa1,Sorokin:2004ie}
 that the equations of motion in the compensator formalism are
invariant for general gauge transformations, but if we try to gauge fix the compensators to zero, constraints on gauge parameters and on gauge fields
appear, namely the gauge parameters must be traceless
and the gauge fields double traceless.
In our framework there are no ways of taking the trace of completely holomorphic tensors,
instead differential constraints appear on gauge parameters and on gauge fields.
To see this, let us use part of the gauge freedom in \eqref{gauge transformation phi} and \eqref{gauge
transformation rho} to make the compensators vanish: $\rho^{ij}=0$. The residual gauge symmetry must satisfy $\bar{Q}^{[i}\ket{\Lambda^{j]}}=0$, and
this can be achieved if the gauge parameters are taken to be
``divergenceless'': $\bar{Q}^i\ket{\Lambda^{j}}=0$ for $i\neq j$.
Similarly, the gauge choice $\rho^{ij}=0$ imposes constraints also on the gauge
field $\phi$.
This can be seen by acting with $\bar{Q}^k$ on both
sides of eq. \eqref{Fronsdal Labastida with compensators} to obtain
\begin{equation}
Q_i\bar{Q}^k\bar{Q}^i\ket{\phi}=Q_i\,\big[\bar{Q}^kQ_j\ket{\rho^{ij}}-H\ket{\rho^{ki}}
\big]\ .
\end{equation}
The right hand side of this equation vanishes in the partially
gauge fixed theory with $\rho^{ij}=0$. For consistency the left hand side must
vanish as well, and this is guaranteed if $\bar{Q}^k\bar{Q}^i\ket{\phi}=0$,
that corresponds to setting to zero all possible double divergences. One may check that
this constraint is kept invariant by gauge transformations with parameters
satisfying $\bar{Q}^i\ket{\Lambda^{j}}=0$ with $i\neq j$.
Once the compensator fields have been eliminated, the gauge potential describing
the particle satisfies the simpler second order wave equation $G\ket{\phi}=0$ that, in tensorial language, reads
\begin{equation}
\partial_\alpha \bar\partial^\alpha \phi_{\mu_1..\,\mu_p,...,\nu_1..\,\nu_p}
-p\;\de_{\mu_1}\bar\de^\alpha\phi_{\alpha\mu_2..\,\mu_p,...,\nu_1..\nu_p}-\ldots
-p\;\de_{\nu_1}\bar\de^\alpha \phi_{\mu_1..\mu_p,...,\alpha\nu_2..\nu_p}=0
\end{equation}
where $p\equiv m-1$, and weighted antisymmetrization is understood on $\mu$'s, $\nu$'s and so on.

In order to clarify the meaning of our quantum mechanical notation, let us analyze in tensorial
language a specific case: $N=2,\,p=2$. This is the simplest model where all of the issues treated so
far appear in a non trivial way. The gauge field $\phi$ has the structure
\begin{equation}
\phi_{\mu_1\mu_2,\nu_1\nu_2}\,\sim\,\begin{array}{c}\\[-15pt]\yng(2,2)\end{array}
\end{equation}
while the unique independent compensator is a symmetric tensor $\rho_{\mu\nu}$. The gauge invariant
equations of motion read
\begin{equation}\label{eom with compensators N=2 p=2}
\de_\alpha\bar\de^\alpha\phi_{\mu_1\mu_2,\nu_1\nu_2}-2\de_{\mu_1}\bar\de^\alpha\phi_{\alpha\mu_2,\nu_1\nu_2}
-2\de_{\nu_1}\bar\de^\alpha\phi_{\mu_1\mu_2,\alpha\nu_2}=2\de_{\mu_1}\de_{\nu_1}\rho_{\mu_2\nu_2}
\end{equation}
with an understood weighted antisymmetrization on the $\mu$ and $\nu$ group of indices,
that will be employed  in all of the following equations as well. 
The gauge transformations for the field $\phi$ and the compensator are given
by
\begin{equation}\label{gauge transformation N=2 p=2}
\delta\phi_{\mu_1\mu_2,\nu_1\nu_2}=\de_{\mu_1}\Lambda_{\nu_1\nu_2,\mu_2}+\de_{\nu_1}\Lambda_{\mu_1\mu_2,\nu_2}
\;,\quad\delta\rho_{\mu\nu}=-\bar\de^\alpha\Lambda_{\alpha\mu,\nu}-\bar\de^\alpha\Lambda_{\alpha\nu,\mu}
\end{equation}
where a factor of $-2i$ has been absorbed in the definition of the gauge parameter, whose Young
tableau is
\begin{equation}
\Lambda_{\mu_1\mu_2,\nu}\,\sim\,\begin{array}{c}\\[-15pt]\yng(2,1)\end{array}\;.
\end{equation}
Using part of the gauge freedom, one can fix the compensator to zero, obtaining the gauge invariant
equation
\begin{equation}\label{eom without compensators N=2 p=2}
\de_\alpha\bar\de^\alpha\phi_{\mu_1\mu_2,\nu_1\nu_2}-2\de_{\mu_1}\bar\de^\alpha\phi_{\alpha\mu_2,\nu_1\nu_2}
-2\de_{\nu_1}\bar\de^\alpha\phi_{\mu_1\mu_2,\alpha\nu_2}=0
\end{equation}
which is left invariant by the gauge transformations \eqref{gauge transformation N=2 p=2} with
constrained gauge parameters
\begin{equation}
\bar\de^\alpha\Lambda_{\alpha\mu,\nu}=0\;.
\end{equation}
For consistency, the gauge field appearing in this equation
must also satisfy a differential constraint
\begin{equation}
\bar\de^\alpha\bar\de^\beta\phi_{\alpha\mu,\beta\nu}=0 
\end{equation}
which is preserved by the gauge transformations with constrained gauge parameters.

To count the physical degrees of freedom, one has to use the remaining gauge freedom to eliminate
unphysical ``polarizations'' from $\phi$. This way one ends up with a gauge field
$\phi_{m_1..m_p,...,n_1..n_p}$, where indices run over $d-2$ directions, \emph{i.e.}
$m,n=1,2,...,d-2$. Perhaps this is best seen in the particle language, since by using the complex
$N$ supersymmetries one can eliminate the fermionic fields $\psi^\mu_i$ and their complex
conjugates with the index $\mu$ pointing along two chosen directions. In this ``light cone gauge''
the tensor $\phi_{m_1..m_p,...,n_1..n_p}$ describes an irreducible representation of the little
group for massless particles, $U(d-2)$, with the same Young tableau of eq. \eqref{phi}. The
dimension of such representation corresponds to the number of physical degrees of freedom of the
particle. Using the ``factors over hook'' rule it is easy to compute the dimension of this Young
tableau, and the resulting degrees of freedom, for all $d$, $N$ and $p$, are
\begin{equation}\label{Dof Young tableau}
Dof(d,N,p)=\prod_{j=0}^{N-1}\frac{j!(j+d-2)!}{(j+p)!(j+d-2-p)!}
\end{equation}
where we recall that $p=m-1=\frac{d}{2}+s-1$. We note that in the case of an odd number of complex
dimensions the physical spectrum is empty unless the Chern-Simons term is added, \emph{i.e.}
$s\neq0$. The quantization of this Chern-Simons coupling
can be understood also from the requirement of
cancelling gauge anomalies \cite{Elitzur:1985xj}.

Let us analyze a few examples.
From \eqref{Dof Young tableau} one can see that in $d=2$ (four \emph{real}
dimensions) without Chern-Simons coupling, there is always one degree of freedom
for any value of $N$:
$Dof(2,N,0)=1$. So, with $s=0$, all the $U(N)$ spinning particle theories propagate only
a scalar field in two complex dimensions, and share for this aspect the features of $N=2$ superstrings,
where only the scalar ground states survive at the critical dimension $d=2$,
see for example the review \cite{Marcus:1992wi}. Another
simple case is the $N=1$ theory in arbitrary complex dimensions:
the field strengths are $(p+1,0)$-forms
$F_{\mu_1..\mu_{p+1}}$, the gauge potentials are $(p,0)$-forms $\phi_{\mu_1..\mu_{p}}$
and \eqref{Dof Young tableau} gives $Dof(d,1,p)=\binom{d-2}{p}$; that is
the number of independent components of an  antisymmetric tensor of $U(d-2)$ with $p$ indices, $\phi_{m_1..m_p}$. In the last section we will compute the
one-loop partition function for the $U(N)$ spinning particle.
After covariantly gauge fixing the action \eqref{Sinvariantconfspace} on the torus, the
path integral reduces to an integral over a corresponding moduli space which
computes the number of physical degrees of freedom. Indeed,
we shall see that they coincide with the canonical computation just presented.

To summarize, we have described gauge invariant equations with compensators
\begin{equation}
G\ket{\phi} = Q_i Q_j\ket{\rho^{ij}}
\end{equation}
with $G=-H +Q_i\bar Q^i$, and gauge symmetries given by
 \begin{equation}
\delta\ket{\phi} = Q_i \ket{\Lambda^{i}} \ , \qquad \delta\ket{\rho^{ij}} =-
\bar{Q}^{[i}\ket{\Lambda^{j]}}
\end{equation}
where $\ket{\rho^{ij}}\equiv \bar V^i \bar V^j \ket{\rho}$,
$\ket{\Lambda^{i}}  \equiv \bar W^i \ket{\Lambda}$ and with
$\ket{\phi}$, $\ket{\rho}$, $\ket{\Lambda}$ describing tensors with rectangular
$p\times N$ Young tableaux of $U(d)$, as in \eqref{phi}.

Similarly, gauge invariant equations without compensators are given by
\begin{equation}
G\ket{\phi} = 0
\end{equation}
with gauge symmetry
 \begin{equation}
\delta\ket{\phi} = Q_i \ket{\Lambda^{i}}
\end{equation}
where  $\ket{\Lambda^{i}}  \equiv \bar W^i \ket{\Lambda}$, with fields and gauge
parameters satisfying the differential constraints
 \begin{equation}
\bar{Q}^i\bar{Q}^j\ket{\phi}=0 \ , \qquad
\bar{Q}^{i}\ket{\Lambda^{j}}=0 \qquad (i\neq j) .
\end{equation}

\section{Supersymmetry algebra in curved K\"ahler manifolds}

We now turn to study the supersymmetry algebra on arbitrary K\"ahler manifolds. It will be
shown that for all $N$ it is possible to close the algebra, though quadratically, on K\"ahler manifolds of constant holomorphic curvature,
and so even for $N>2$ a consistent quantization can be
obtained beyond the case of flat space.

Looking at the quantum algebra \eqref{quantum constraints}, we note that
the last three relations
just state that $J_i^j$ are $U(N)$ generators and that $Q_i,\bar{Q}^j$ belong to the $\mathbf{N}$, $\mathbf{\bar{N}}$ of $U(N)$, and presumably
these relations should be left unchanged even in curved space. The first equation is
the key ingredient of the supersymmetry algebra, and is going to be modified
by a nonvanishing curvature. Our aim is to deform the algebra
\eqref{quantum constraints} introducing curvature, but keeping it first
class, as necessary if we want to impose the corresponding constraints consistently.

Thus, let us consider the theory on an arbitrary K\"ahler manifold.
 The only non vanishing components of the
metric are $g_{\mu\bar{\nu}}(x,\bar x)=g_{\bar{\nu}\mu}(x,\bar x)$, which lead to
nonvanishing Christoffel
coefficients for the total holomorphic or antiholomorphic parts only:
$\Gamma^\mu_{\nu\lambda}$, $\Gamma^{\bar{\mu}}_{\bar{\nu}\bar{\lambda}}$.
In curved space we will
use fermions with flat indices: $\psi^a_i$ and $\bar{\psi}^{\bar{a} i}$;
the $U(N)$ generators are essentially unchanged, being defined by
$J_i^j=\frac{1}{2}\big[\psi^a_i\,,\bar{\psi}_a^j\big]$ (the flat tangent
metric is simply $\delta_{a\bar{b}}$), but the supercharges need a suitable covariantization.
Since the holonomy group of K\"ahler manifolds of
real dimension $D=2d$ is $U(d)$, the connection would be a $U(d)$ spin connection, and the
covariant derivative reads
\begin{equation*}
\nabla_\mu=\de_\mu+\omega_{\mu a\bar{b}}M^{a\bar{b}}
\end{equation*}
where $M^{a\bar{b}}$ are the $U(d)$ generators. In the particle model
these generators can be realized by
\begin{equation}\label{Mwithsymmetricordering}
M^{a\bar
b}=\frac{1}{2}\big[\psi^a_i\,,\bar{\psi}^{\bar{b}\,i}\big]=\psi^a_i\bar{\psi}^{\bar{b}i}-\frac{N}{2}\delta^{a\bar{b}}
\end{equation}
as they satisfy indeed the Lie algebra of $U(d)$
\begin{equation*}
\big[M^{a\bar{b}}\,,M^{c\bar{d}}\big]=\delta^{c\bar{b}}M^{a\bar{d}}
-\delta^{a\bar{d}}M^{c\bar{b}}\;.
\end{equation*}
In this way we construct covariantized momenta\footnote{We denote
$g=\text{det} g_{\mu\bar\nu}$ and the $g$ factors ensure hermiticity.}
\begin{equation} \label{covmom}
\begin{split}
\pi_\mu &= g^{1/2} \big(p_\mu-i\omega_{\mu a\bar{b}}M^{a\bar{b}}\big)g^{-1/2} \\
\bar \pi_{\bar{\mu}} &=
g^{1/2}\big(\bar{p}_{\bar{\mu}}-i\omega_{\bar{\mu} a\bar{b}}M^{a\bar{b}}\big)g^{-1/2}\;,
\end{split}
\end{equation}
and supercharges
\begin{equation}
Q_i=\psi^a_i\,e_a^\mu\,\pi_\mu \quad,\quad
\bar{Q}^j=\bar{\psi}^{\bar{a} j}\,e_{\bar{a}}^{\bar{\mu}}\,\bar \pi_{\bar{\mu}}\;.
\end{equation}
With these charges the $JJ$, $JQ$ and $J\bar{Q}$ commutators are the same as before, but the
$Q\bar{Q}$ anticommutator now reads
\begin{equation}\label{QQbar}
\big\{Q_i\,,\bar{Q}^j\big\}=\delta_i^j\,H_0-R_{a\bar{b}c\bar{d}}\,\psi_i^a\bar{\psi}^{\bar{b}j} M^{c\bar{d}}\;,
\end{equation}
where $H_0=g^{\bar{\mu}\nu}\bar\pi_{\bar{\mu}}\pi_\nu$ is the minimal covariantization of the
hamiltonian. As in the case of $O(N)$ supersymmetry
\cite{Kuzenko:1995mg,Bastianelli:2008nm},
we can achieve the closure of the algebra on
particular manifolds, namely K\"ahler manifolds with constant holomorphic
 curvature, which admit a Riemann tensor of the form \cite{goldberg}
\begin{equation}\label{constholomcurv}
R_{a\bar{b}c\bar{d}}=\Lambda\,\big(\delta_{a\bar{b}}\delta_{c\bar{d}}+\delta_{a\bar{d}}\delta_{c\bar{b}}\big)\;,
\end{equation}
with constant $\Lambda$. As for real manifolds maximally symmetric spacetimes
are de Sitter, anti-de Sitter and flat Minkowski space,
prototypes of K\"ahler manifolds with a Riemann tensor of the form
\eqref{constholomcurv} are the complex projective space $\mathbb{C}\mathbb{P}^d$, complex
hyperbolic space $\mathbb{C}\mathbb{H}^d$ and, of course, flat complex space $\mathbb{C}^d$ viewed as a K\"ahler manifold. Inserting the $U(d)$ generators
$M^{a\bar b}=\frac{1}{2}\big[\psi^a_i\,,\bar{\psi}^{\bar{b}\,i}\big]$,
the $\big\{Q,\bar{Q}\big\}$ anticommutator closes quadratically
(up to an obvious redefinition of the hamiltonian)
\begin{equation}
\big\{Q_i\,,\bar{Q}^j\big\}=\delta_i^j\,\big(H_0-a\,J-b\big)-\Lambda J_i^j\,
J+\frac{\Lambda}{2}\,\big\{J_i^k\,,J_k^j\big\}\;,
\end{equation}
with $J=J_k^k$, $a=\Lambda\frac{d+1}{2}$ and $b=\Lambda\frac{d(N+d)}{4}$. The hamiltonian $H_0$
has, however, an unusual commutator with the supercharges, namely
\begin{equation}
\begin{split}
\big[H_0\,,Q_i\big] &= -\Lambda\, J_i^k\,Q_k+\Lambda\, J\,Q_i+\Lambda\frac{N+d}{2}\,Q_i\quad,\\
\big[H_0\,,\bar{Q}^i\big] &= -\big[H_0\,,Q_i\big]^\dagger\;,
\end{split}
\end{equation}
so we add to $H_0$ a hermitian and $U(N)$ neutral $J$ combination in order to cancel the commutators above.
We recall that, including a Chern-Simons coupling, the quantum constraint on $J$ is
$J_i^j-s\delta_i^j=0$ and so, in order to make manifest the quadratic closure of our algebra, we set $\tilde J_i^j=J_i^j-s\delta_i^j$ and $\tilde J=\tilde J_i^i$,
finally obtaining
\begin{equation}\label{quadraticalgebra}
\begin{split}
\big[H\,,\tilde J_i^j\big] &= \big[H\,,Q_i\big]=\big[H\,,\bar{Q}^j\big]=0 \\
\big[\tilde J_i^j\,,\tilde J_k^l\big] &= \delta_k^j\, \tilde J_i^l-\delta_i^l\, \tilde J_k^j \\
\big[\tilde J_i^j\,,Q_k\big] &= \delta_j^k\,Q_i\quad,\quad
\big[\tilde J_i^j\,,\bar{Q}^k\big]=-\delta_i^k\,\bar{Q}^j \\
\big\{Q_i\,,\bar{Q}^j\big\} &= \delta_i^jH+\Lambda\left[\tilde J_i^k\,\tilde J_k^j-\tilde
J_i^j\tilde J+h_1\tilde J_i^j+\frac{1}{2}\delta_i^j\left(\tilde J^2-\tilde J_k^l\tilde
J_l^k+h_2\tilde J\right)\right]\;,
\end{split}
\end{equation}
where the complete hamiltonian reads
\begin{equation} \label{ham}
H=H_0+\frac{\Lambda}{2}\left[J_i^kJ_k^i-J^2-h_3J-h_4\right]\;,
\end{equation}
with the $h_i$ being defined by
\begin{equation}
\begin{split}
h_1 &= \big (2-N \big)s-\frac{N}{2}\;, \\
h_2 &= 2s\big (N-2\big )+1\;,\qquad h_3=d+1\;, \\
h_4 &= \frac{d}{2}\,\big(N+d\big)-s^2\big(N-1\big)\big(N-2\big)\;.
\end{split}
\end{equation}
 This is no more a Lie algebra but, being still first class, permits a consistent
realization of the constraints $\tilde J_i^j=H=Q_i=\bar{Q}^j=0$,
which define higher spin equations on such curved backgrounds. As the
analogous result obtained in \cite{Bastianelli:2008nm} for the $O(N)$ spinning particle, the
quadratic algebra \eqref{quadraticalgebra} coincides with the zero mode, in the Ramond sector, of the quadratic
$U(N)$ superconformal algebra found by Bershadsky and Knizhnik in \cite{Bershadsky:1986ms,Knizhnik:1986wc}.

Up to now we have used $U(d)$ generators with the preferred ordering given in
\eqref{Mwithsymmetricordering}, but a quadratic closure of the supersymmetry algebra can be
achieved with an arbitrary ordering,
corresponding to a different coupling to the $U(1)$ part of the spin connection
$\omega_\mu=\omega_{\mu a\bar{b}}\delta^{a\bar{b}}$: if
in eq. \eqref{covmom} we choose as $U(d)$ generators
\begin{equation}
\mathcal{M}^{a\bar{b}}=\psi^a_i\bar{\psi}^{\bar{b}i}-c\delta^{a\bar{b}}\;,
\end{equation}
with arbitrary $c$, \eqref{QQbar} remains unchanged in form,
and choosing the Riemann tensor as in \eqref{constholomcurv},
the quadratic algebra in \eqref{quadraticalgebra} and \eqref{ham} maintains
the same structure but with different numerical
coefficients $h_i\to h_i(c)$, given by
\begin{equation}
\begin{split}
h_1(c) &= \big(2-N \big )s-\frac{d}{2}\,\big(N-2c\big)+c-N \\
h_2(c) &= \big(d+1\big)\big(N-2c\big)+2s\big(N-2\big)+1\\
h_3(c) &= \big(d+1\big)\big(N-2c+1\big)\\
h_4(c) &=
d\left[\frac{d}{2}\big(N-2c+1\big)+N-c\right]+s\big(N-1\big)
\left[ \big(d+1\big)\big(2c-N\big)-s\big(N-2\big) \right ] \ .
\end{split}
\end{equation}
To recover the previous results is sufficient to put $c=N/2$ in the above formulas.

With this constraint algebra at hand it is possible to achieve the quantization of the $U(N)$
particle, for all $N$, on K\"ahler manifolds of constant holomorphic  curvature.
We expect that a consistent quantization can be achieved also on more general K\"ahler
manifold, namely those possessing a vanishing Bochner tensor,
a K\"ahler analogue of the conformal Weyl tensor,
but we have not worked out the explicit constraint algebra.

\section{Partition function and degrees of freedom}

In order to extract from the $U(N)$ spinning particle action \eqref{Sinvariantconfspace}
the number of physical excitations, we proceed in computing the one-loop partition
function that gives,
as its first Seeley-DeWitt coefficient, the number of degrees of freedom.
Of course, other heat kernel coefficients vanish in flat space, but once the
measure over the moduli space arising from the gauge fixing procedure
is correctly identified,
one could perform, in principle, more general path integral calculations
to investigate the quantum properties of the field equations on the
backgrounds described previously.

In order to deal with gaussian path integrals rather than oscillating ones, we perform as usual a
Wick rotation on the proper time $\tau\to-i\tau$ and on the gauge field $a^i_j\to ia^i_j$.
The resulting euclidean action reads
\begin{equation}
S[X,G]=\int_0^1d\tau\,\left[e^{-1}\big(\dot{x}^\mu-\bar{\chi}^i\psi_i^\mu\big)\big(\dot{\bar{x}}_\mu-\chi_j\bar{\psi}^j_\mu\big)
+\bar{\psi}_\mu^i\big(\delta_i^j\de_\tau-ia^j_i\big)\psi^\mu_i-isa_i^i\right]
\end{equation}
and is invariant under the supergravity transformations in euclidean time
\begin{equation}\label{gaugetransf euclidian}
\begin{split}
\delta e &= \dot{\xi}+\bar{\chi}^i\epsilon_i+\chi_i\bar{\epsilon}^i\\
\delta\chi_i &=
\dot{\epsilon}_i-ia^k_i\epsilon_k+i\alpha^k_i\chi_k\\
\delta\bar{\chi}^i &= \dot{\bar{\epsilon}}^i+ia^i_k\bar{\epsilon}^k-i\alpha^i_k\bar{\chi}^k\\
\delta a^i_j &= \dot{\alpha}^i_j -i a_j^k\alpha_k^i+ia^i_k\alpha^k_j \ .
\end{split}
\end{equation}
The partition function is obtained by performing the functional integral on a circle,
taking periodic boundary conditions for the bosonic fields, and antiperiodic ones
for the fermionic fields
\begin{equation}
Z=\int_{S^1}\frac{DXDG}{\text{Vol(Gauge)}}\,e^{-S[X,G]}
\end{equation}
where, in condensed notation, $X \equiv (x,\bar{x},\psi,\bar{\psi})$ refers to the
matter fields, while $G \equiv (e,\chi,\bar{\chi},a)$ represents the supergravity
multiplet. Since our model is a gauge theory, it is necessary to divide by the volume
of the gauge group. The
gauge fixing procedure can be achieved with the standard Faddeev-Popov method.
We select a covariant gauge by imposing gauge fixing conditions on the worldline
supergravity fields. The latter can be gauged away, except
for a remaining finite number of modular integrations that take into account gauge
inequivalent configurations.
We follow the same strategy employed in \cite{Bastianelli:2007pv}
for the $O(N)$ spinning particle, to which we refer for additional details.\\

\textbf{Gauge fixing on the circle}\\
The einbein $e(\tau)$ has periodic boundary conditions and is characterized by the gauge
invariant quantity $\beta=\int_0^1e(\tau)d\tau$, which represents the invariant length of
the circle. A standard gauge for worldline reparametrizations is to fix
$e(\tau)=\beta$, and the path integral over $e$ reduces to an
ordinary integral over the usual proper time $\beta$,
with the familiar ``one-loop'' measure
\begin{equation*}
\int_0^{\infty}\frac{d\beta}{\beta}\;.
\end{equation*}
Due to antiperiodic boundary conditions,
the complex gravitini $\chi_i$ and $\bar{\chi}^i$ can
be completely gauged away, $\chi_i(\tau)=\bar{\chi}^i(\tau)=0$,
leaving corresponding  Faddeev-Popov determinants of the differential operators
that can be extracted from \eqref{gaugetransf euclidian}.
Finally, the gauge field
$a_i^j$ can have nontrivial Wilson loops around the circle,
that capture the complete gauge invariant information contained in them.
They can be gauge fixed to a constant
hermitian $N\times N$ matrix, $a_i^j(\tau)=\theta_i^j$, that can be always diagonalized
through a constant $U(N)$ gauge transformation
\begin{equation}\label{A gauge fixed}
\theta_i^j\,\rightarrow\;\left(
\begin{array}{ccc}
\theta_1 & {} & {}\\
{} & \ddots & {}\\
{} & {} & \theta_N
\end{array}\right)\;.
\end{equation}
Recalling that $a_i^j$ belongs to the Lie algebra of $U(N)$, we see by exponentiation that the
$\theta_i$ are in fact angles ranging from $0$ to $2\pi$. Now, the path integral over $x$ and
$\bar{x}$ gives as usual $V\,(2\pi \beta)^{-d}$, where
$V=i^d\int d^dx_0d^d\bar{x}_0$ (the integral over the $x$ zero modes)
is the spacetime volume. The $D\psi D\bar{\psi}$ integral gives
$\text{Det}_{A}(\delta_i^j\de_{\tau}-i\theta_i^j)^d$,
while integrals over the susy ghosts give a power $-2$ of the same determinant.
Subscripts $P$ and $A$  keep track of the periodic or antiperiodic
boundary conditions.
From the diagonalization  \eqref{A gauge fixed}, we see that
the integration over the moduli space of $a_i^j$ reduces to integration over the angles
\begin{equation}
\frac{1}{N!}\prod_{i=1}^N\int_0^{2\pi}\frac{d\theta_i}{2\pi}\;,
\end{equation}
and division by $N!$ is needed to eliminate the overcounting due to the permutations of
the $\theta$'s, that are all gauge equivalent.
The last integration to be performed is over the ghosts for the gauge group $U(N)$,
that gives $\text{Det}'_{P}(\de_\tau+i\theta_{\text{adj}})$,
\emph{i.e.} with the zero
modes removed and the gauge fixed $a_i^j$ taken in the adjoint representation,
as follows from
$\delta a_i^k=\mathcal{D}\alpha_i^k$ in \eqref{gaugetransf euclidian}.
Now, we use the diagonalized form \eqref{A gauge fixed}, and putting together the
various contributions we obtain for the partition function
\begin{equation}
\begin{split}
Z\propto & \,V\int_0^{\infty}\frac{d\beta}{\beta} \frac{1}{(2\pi \beta)^d}
\frac{1}{N!}\prod_{i=1}^N\int_0^{2\pi}\frac{d\theta_i}{2\pi}\;
e^{-is\theta_i}
\text{Det}_{A}(\de_\tau-i\theta_i)^{d-2}\\
& \times\prod_{k\neq l}\text{Det}_{P}(\de_\tau-i(\theta_k-\theta_l))\;.
\end{split}
\end{equation}
These determinants are standard ones and can be computed using operator methods with simple
fermionic systems. Namely, they are:
$\text{Det}_{A}(\de_{\tau}- i\theta)=2\cos\frac{\theta}{2}$ and
$\text{Det}_{P}(\de_\tau-i\theta)=2i\sin\frac{\theta}{2}$. Substituting in the expression for $Z$ one
finally finds
\begin{equation}
\begin{split}
Z\propto & \,V\int_0^{\infty}\frac{d\beta}{\beta}\frac{1}{(2\pi \beta)^d}\,
\left[\frac{1}{N!}\prod_{i=1}^N\int_0^{2\pi}
\frac{d\theta_i}{2\pi}\; e^{-is\theta_i}
\left(2\cos\frac{\theta_i}{2}\right)^{d-2}
\prod_{k<l}\left(2\sin\frac{\theta_k-\theta_l}{2}\right)^2\right]\;.
\end{split}
\end{equation}\\[-5mm]

\textbf{Degrees of freedom}\\
The part in square brackets of the above formula gives the number of degrees of
freedom of the particle, since the rest is simply the partition function for the center
of mass, and so we have the following expression for the physical degrees of freedom
\begin{equation}\label{Dof PI integral}
Dof(d,N;s)=\frac{1}{N!}\prod_{i=1}^N\int_0^{2\pi}\frac{d\theta_i}{2\pi}\;
e^{-is\theta_i}
\left(2\cos\frac{\theta_i}{2}\right)^{d-2}
\prod_{k<l}\left(2\sin\frac{\theta_k-\theta_l}{2}\right)^2\;.
\end{equation}
It is normalized to $Dof(d,0;0)=1$ for $N=0$, which corresponds to
a simple scalar field.
It is now convenient to go to complex coordinates: $z_i=e^{i\theta_i}$. Recalling that
$s=m-\frac{d}{2}=p+1-\frac{d}{2}$, the above expression in terms of $p$ becomes
\begin{equation}\label{Dof PI z integral}
Dof(d,N,p)=\frac{1}{N!}\prod_{i=1}^N\oint\frac{dz_i}{2\pi
i}\frac{1}{z_i^{p+1}}(z_i+1)^{d-2}
\prod_{k<l}\ass{z_k-z_l}^2
\end{equation}
where the integration contour is the unit circle around the origin in $\mathbb{C}$, $\forall  i$.
Now, we perform a new change of variables, passing from the unit complex circle to the
real line by means of stereographic projection: $z_j=\frac{i-x_j}{i+x_j}$.
The integral becomes
\begin{equation}\label{Dof final}
Dof(d,N,p)=\frac{2^{N^2+Nd-3N}}{N!\pi^N}\,\int_{\mathbb{R}^N}d^Nx\;
\ass{\Delta(x)}^2\prod_{j=1}^N
(1+ix_j)^{-(N+p)}(1-ix_j)^{-(d+N-p-2)}
\end{equation}
where we have recognized the square of the Van der Monde determinant
\begin{equation}
\Delta(x)=\prod_{i<j}(x_i-x_j)\;.
\end{equation}
Written in term of the $x_i$ variables, \eqref{Dof final} is seen to belong to
a wide class of Selberg's integrals, that can be computed by means of orthogonal polynomials techniques\footnote{Much information and many details about these techniques can be found
in \cite{Mehta}.}.
The known Selberg's integral in question, that can be found in \cite{Mehta}, reads
\begin{equation}\label{Selberg integral}
\begin{split}
& J(a,b,\alpha,\beta,\gamma,n)=
\int_{\mathbb{R}^N}d^Nx\;
\ass{\Delta(x)}^{2\gamma}\prod_{j=1}^n(a+ix_j)^{-\alpha}(b-ix_j)^{-\beta}\\
&= \frac{(2\pi)^n}{(a+b)^{(\alpha+\beta)n-\gamma n(n-1)-n}}
\prod_{j=0}^{n-1}\frac{\Gamma(1+\gamma+j\gamma)\Gamma(\alpha+\beta-(n+j-1)\gamma-1)}
{\Gamma(1+\gamma)\Gamma(\alpha-j\gamma)\Gamma(\beta-j\gamma)}
\end{split}
\end{equation}
valid for Re$a$, Re$b$, Re$\alpha$, Re$\beta>0$, Re$(\alpha+\beta)>1$, and
\begin{equation*}
-\frac{1}{n}<\text{Re}\gamma<\min\left(\frac{\text{Re}\alpha}{n-1},\,\frac{\text{Re}\beta}{n-1},\,\frac{\text{Re}(\alpha+\beta+1)}{2(n-1)}\right)\;.
\end{equation*}
Our eq. \eqref{Dof final} corresponds to this form of the Selberg's integral with
$(a=b=\gamma=1,\,\alpha=N+p,\,\beta=d+N-p-2,\,n=N)$ so, with \eqref{Selberg integral}
at hand, after a little algebra, we obtain the final result
\begin{equation}
\begin{split}
Dof(d,N,p) &=
\frac{2^{N^2+Nd-3N}}{N!\pi^N}\,J(1,1,N+p,d+N-p-2,1,N)\\
&=  \prod_{j=0}^{N-1}\frac{j!(j+d-2)!}{(j+p)!(j+d-2-p)!}
\end{split}
\end{equation}
that agrees with the dimension of the rectangular Young tableau of
$U(d-2)$ with $p$ rows and $N$ columns, as in \eqref{Dof Young tableau},
thus reproducing the number of physical polarizations predicted by canonical
quantization.

\section{Conclusions and outlook}

We have analyzed $U(N)$ spinning particles and obtained from them new gauge invariant
higher spin equations that live on complex spaces.
These equations define a complex version of the standard higher spin equations
of Minkowski spacetime \cite{Fronsdal:1978rb,Labastida:1986ft,Sorokin:2004ie}.
We have obtained them by integrating a subset of the constraints that
arise form the Dirac quantization of the $U(N)$ spinning particle.
The spinning particle language is quite efficient, as already
exemplified in \cite{Bastianelli:2008nm} for the $O(N)$ spinning particle, in which case
it allowed to describe in a simple way the structure of minkowskian
higher spin fields, including the use of compensators \cite{Francia:2002aa1}
and the application of generalized Poincar\'e lemmas
to integrate higher order field equations
\cite{DuboisViolette:2001jk,Bekaert:2002dt,de Medeiros:2002ge,Bandos:2005mb}.
Similar constructions have been presented here for the new class
of complex higher spin equations. Having described these equation
on a flat complex manifold, we have shown in principle their consistency
also on a more general class of K\"ahler manifolds, namely those
K\"ahler manifolds with constant holomorphic  curvature,
as in this case the algebra of the quantum constraints closes in a quadratic way
and remains first class. An important feature of this algebra
is that it realizes in a geometrical way the zero mode sector of the nonlinear
two dimensional $U(N)$ superconformal algebras, introduced
sometimes ago by Bershadsky and Knizhnik \cite{Bershadsky:1986ms,Knizhnik:1986wc}.
Finally we have considered the path integral quantization on the circle
of the $U(N)$ spinning particle in flat space, corresponding to the
one-loop effective action of the quantized version of the higher spin equations
introduced earlier.
This way we have calculated the number of physical degrees of freedom for all $d$, $N$
and $p$, and checked the correctness of our path integral construction containing,
in particular, the measure on the moduli space of the $U(N)$ extended supergravity
on the circle.

As for future developments,
an application of this worldline approach could be to compute
perturbatively the one-loop effective action on arbitrary K\"ahler manifolds
for the $U(1)$ and $U(2)$ models, as done for the similar cases of the
$O(N)$ spinning particle on arbitrarily curved spaces with $N=0,1,2$,
which produced the effective action
for scalars \cite{Bastianelli:2002fv}, spin 1/2 \cite{Bastianelli:2002qw},
and arbitrary differential forms (including vectors) \cite{Bastianelli:2005vk}
coupled to gravity, respectively.
Similarly, one could consider the $U(N)$  models with $N>2$ on
K\"ahler manifolds with constant holomorphic  curvature
and compute the corresponding partition function on the circle (\emph{i.e.}
the one loop effective action of the corresponding quantum field theory).
Finally, it could be interesting to study along similar lines
gauged versions of various quantum mechanical models,
for example those described in
\cite{FigueroaO'Farrill:1997ks,Zucchini:2006ds,Hallowell:2007qk,Burkart:2008bq},
to unearth novel gauge invariant field equations.

\acknowledgments{
This work was supported in part by the Italian MIUR-PRIN contract 20075ATT78.}
\vskip2cm

\end{document}